\documentclass[aps,twocolumn,showpacs,preprintnumbers]{revtex4-1}

\usepackage{psfrag,graphicx}
\usepackage{dcolumn}
\usepackage{amsmath,amssymb}
\usepackage{bm}
\usepackage{amsfonts,amssymb,amsmath}        
\usepackage{epstopdf}

\newcommand{\be}{\begin{equation}}
\newcommand{\ee}{\end{equation}}
\newcommand{\bq}{\begin{eqnarray}}
\newcommand{\eq}{\end{eqnarray}}

\newcommand{\ket}[1]{\left | \, #1 \right\rangle}
\newcommand{\bra}[1]{\left \langle #1 \, \right |}
\begin{document}

\title{Towards unambiguous calculation of the topological entropy for mixed states}
\date{\today}
\author{James R. Wootton}
\affiliation{School of Physics and Astronomy, University of Leeds, Leeds LS2 9JT,
U.K.}

\begin{abstract}

Calculation of topological order parameters, such as the topological entropy and topological mutual information, are used to determine whether states possess topological order. Their calculation is expected to give reliable results when the ground states of gapped Hamiltonians are considered, since non-topological correlations are suppressed by a finite correlation length. However, studies of thermal states and the effects of incoherent errors require calculations involving mixed states. Here we show that such mixed states can effectively lead to a diverging correlation length, and hence may give misleading results when these order parameters are calculated. To solve this problem, we propose a novel method to calculate the quantity, allowing topologically ordered states to be identified with greater confidence.

\end{abstract}


\maketitle


Topologically ordered states have generated a great deal of interest, both in condensed matter physics \cite{wen,honeycomb,stringnet,hamma1,hamma2} and quantum information theory. This is due in the most part to their ability to support anyonic quasiparticles, useful for fault-tolerant quantum computation \cite{preskill,dennis,freedman,pachos,wootton}. Topologically ordered states cannot, in general, be detected by any local order parameter. Instead order parameters such as the topological entropy and topological mutual information must be used \cite{kitpres,levwen,iblisdir}. These aim to detect the unique non-local correlations found in topologically ordered states. However, present formulations are known to also have a contribution from non-topological correlations \cite{kitpres,levwen}. When pure ground states of gapped Hamiltonians are considered, these non-topological correlations only extend for a short range determined by their correlation length. Their contributions are therefore expected to be suppressed when the regions used to evaluate the entropies are made larger than this length scale. However, here we show that statistical mixtures of the degenerate ground states of gapped Hamiltonians with local interactions can effectively have a diverging correlation length, preventing the suppression of non-topological contributions. This could cause non-topological states to be identified as topological, or topological states to be identified as non-topological. This is an important matter to address, since calculations using mixed states are required for the studies of thermal states and incoherent error models that are of high physical and practical significance \cite{claudios1,claudios2,iblisdir,melko}. To solve the problem, a modified definition of the topological entropy is proposed. This correctly identifies and discards the non-topological contributions, and so can be used with confidence the determine whether topological order is present for mixed states.

To begin, it will be advantageous to introduce the concepts which will be used throughout this work. The set of anyonic quasiparticles supported by a given topologically ordered state can be described in terms of an anyon model. Within each of these models the quantum dimension $d_a$ can be defined for each species of anyon, $a$ \cite{preskill,pachos}. The total quantum dimension of the model is then given by $D^2 =\sum_a d_a^2$. The case of $D=1$ corresponds to the trivial anyon model, which contains only the vacuum sector, and is associated with non-topological states. The case of $D>1$, however, is a signature of topologically ordered states.

The realization of topological order requires large many-body systems. We use $\rho$ to denote the state of such a system and $\rho_R = {\rm tr}_{R_c} \rho$ to denote the state of the subsystem contained within an arbitrary region $R$, where $R_c$ is its complement. Using the Von-Neumann entropies of the states of these regions (for example $S_R = -{\rm tr} (\rho_R \ln \rho_R)$) the mutual information between two regions $R$ and $R'$ may be defined as $I_{R,R'} = S_R + S_{R'} - S_{RR'}$ \cite{ikemike}. Here $RR'$ is used as shorthand for the composite region $R \cup R'$. We will use the convention that, if only a single region is specified in the subscript, the mutual information is between that region and it's complement, e.g. $I_R \equiv I_{R,R_c}$.

\section{The topological order parameters} 

The two main methods to calculate a topological order parameter for mixed states are the topological entropy of Levin and Wen \cite{levwen} (and a variant by Castelnovo and Chamon \cite{claudios1}) and the topological mutual information of Iblisdir et al. \cite{iblisdir} (based on the pure state topological entanglement entropy of Kitaev and Preskill \cite{kitpres}). Both of these measures result in the same value of $2\log D$ for topologically ordered states associated with an anyon model of total quantum dimension $D$.

\begin{figure}[t]
\begin{center}
{\includegraphics[width=8cm]{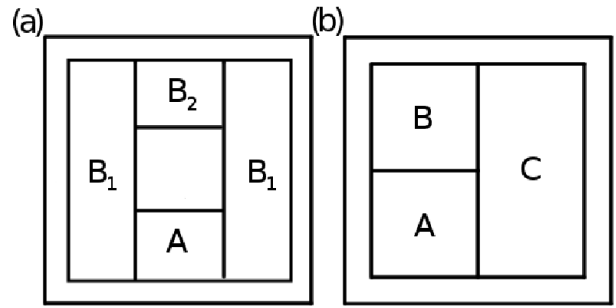}}
\caption{\label{fig1} The regions used for the calculation of a: (a) the topological entropy in Eq. (\ref{LW}); (b) the topological mutual information in Eq. (\ref{KP}). For both of these entropy calculations, the size of the regions should be taken to be large for the most accurate results.}
\end{center}
\end{figure}

The method of Levin and Wen notes that states of topologically ordered systems have correlations for regions that form closed loops that do not exist for those along open loops \cite{levwen}. As such an entropic difference between the two cases is expected. This can be measured by taking an annulus and dividing it into two regions, $A$ and $B$, then further dividing $B$ into $B_1$ and $B_2$, as shown in Fig. \ref{fig1}(a). Using these,
\bq \label{LW0}
\gamma_{LW} + \ldots &=& (S_{B} - S_{B_1})-(S_{AB} - S_{AB_1}).
\eq
Here the ellipsis represents contributions from non-topological correlations, which will be suppressed as the regions are made larger than the correlation length. In this work, we note that this calculation of the topological entropy may alternatively be expressed in terms of mutual informations as,
\be \label{LW}
\gamma_{LW} + \ldots = I_{A,B} - I_{A,B_1}.
\ee
Note that this is not a modification of the quantity, just an alternative way of expressing it. By expressing it in this way we see that it is the information that one section of a closed string ($A$) shares with the rest ($B$), but does not share with its neighbouring sections ($B_1$). This is because $A$ is correlated to $B$ by both topological and local correlations, but is correlated to $B_1$ by only local correlations. The difference between these mutual informations thus isolates the topological correlations. There is no reason to suppose that this argument applies only to pure states. The entropy as defined by Levin and Wen is therefore equally applicable to mixed states. As such, this entropy has been applied to mixed states in previous studies \cite{claudios1,claudios2,melko}. It should be noted that a modified definition for mixed states has also been proposed in \cite{claudios1}, but it gives equivalent results and so will not be considered explicitly here.

The topological mutual information of Iblisdir et al. is a generalization of the topological entanglement entropy of Kitaev and Preskill \cite{kitpres} to mixed states. It uses the fact that the mutual information between the region $R$ and its complement, $R_c$, will satisfy an area law of the following form,
\be \nonumber
I_{R} = \alpha L -\gamma_{I} + \ldots.
\ee
Here $\alpha$ is a positive constant and the correction  $\gamma$ is due to topological correlations. The ellipsis represents contributions from non-topological correlations, which disappear as the regions are made larger than the correlation length.

To cancel out the boundary terms and isolate the topological term, the following linear combination of mutual informations is taken using the regions shown in Fig. \ref{fig1}(a),
\bq \nonumber
\gamma_{I} + \ldots &=& - I_{A} - I_{B} - I_{C} - I_{ABC}\\ \label{KP}
&+& I_{AB} + I_{AC} + I_{BC}.
\eq
Note that the quantity called the topological mutual information in \cite{iblisdsir} is actually $-\gamma_I$. However, $\gamma_I$ is used here such that the values of both measures considered are positive for topologically ordered states. With this definition, topologically ordered states associated with an anyon model of total quantum dimension $D$ will have $\gamma_{LW} = \gamma_I = 2 \log 2$, and non-topologically ordered states will have $\gamma_{LW} = \gamma_I = 0$.

It is important to note that $\gamma_{LW}$ and $\gamma_I$ are measures of topological correlations, not of entanglement. For pure states, the distinction is unnecessary, since the presence of topological (or any other) correlations implies that entanglement is present. However, for mixed states, topological correlations may arise from classical probability distributions, as noted previously in \cite{claudios1}. This property can cause some ambiguity, since it is possible to confuse `true' topological order which is quantum in nature with classical topological order. However, it is not this problem that we address here.

\section{Long-range non-topological correlations} 

The corrections present in the calculations above are due to non-topological correlations that may be present in the system. These can be characterized by by two-point correlation functions of the form $G_{i,j} = \langle O_i O_j \rangle - \langle O_i \rangle \langle O_j \rangle$, where $O_i$ is an operator defined in the neighbourhood of a point $i$ of the system. It is known that, for (pure) ground states of gapped Hamiltonians, $G_{i,j} = O(e^{-d/\xi})$, where $d$ is the distance between points $i$ and $j$. As such, these correlations are suppressed beyond the finite correlation length $\xi$. Making all length scales for the regions used in the above calculations much larger than $\xi$ therefore successfully suppresses the corrections. However, difficulties can arise when mixed states are considered. Though $G_{i,j}$ decays with distance for each individual degenerate ground state of a gapped Hamiltonian, it does not necessarily do so for a mixture of them. The mixed state can therefore, effectively, have an infinite correlation length. Such long-range non-topological correlations will then cause the corrections to remain finite for regions of arbitrary size, and the results of the calculations in Eq. (\ref{LW}) and Eq. (\ref{KP}) to give an ambiguous characterization of topological order. 

As we will show below, this effect is not restricted to unrealistic states. Instead, it can occur for ground states of local gapped Hamiltonians. Furthermore, since the equally weighted mixture of degenerate ground states describes the thermal state at zero temperature, it is an important and realistic state to consider for any Hamiltonian. As such, it is vitally important to determine a definition of the topological entropy that is immune to such effects. 

\begin{figure}[t]
\begin{center}
{\includegraphics[width=4cm]{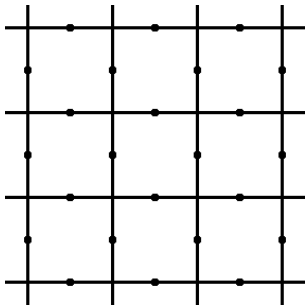}}
\caption{\label{fig2} The spin lattice on which the toric code, and hence the states $\rho^{(1)}$ and $\rho^{(2)}$, is defined. Dots denote spins, and operators are defined around each plaquette and vertex. The lattice has periodic boundary conditions.}
\end{center}
\end{figure}

Two simple examples of states with long-range non-topological correlations can be defined using the toric code model \cite{dennis} with a modified Hamiltonian. This model uses the spin lattice of Fig. \ref{fig2}, defining operators $A_v = \prod_{i \in v} \sigma^x_i$ on the spins around each vertex $v$, and operators $B_p = \prod_{i \in p} \sigma^z_i$ around the spins of each plaquette $p$. The usual Hamiltonian of the model is,
\be \nonumber
H_0 = -J \sum_v A_v -J \sum_p B_p.
\ee
The $A_v$ and $B_p$ operators define the occupation of anyons. States within their $-1$ ($+1$) eigenspace denote that an anyon is present (not present) on the corresponding vertex or plaquette. The Hamiltonian assigns energy to anyons, and hence the ground states of the Hamiltonian are those with no anyons anywhere. Any eigenstate of the Hamiltonian is topologically ordered, with $\gamma_{LW} = \gamma_I = 2 \log 2$. The same is true for any mixture of eigenstates if they all correspond to the same anyon configuration. As such, the ground states of the Hamiltonian have $\gamma_{LW} = \gamma_I = 2 \log 2$, as does any mixture of them. We will use $\rho^{(0)}$ to denote their equally weighted mixture.

For mixtures of $H_0$ eigenstates with different anyon configurations, the topological order parameters can take any value from $0$ to $2 \ln 2$, inclusive. The value $\gamma_{LW} = \gamma_I = 2 \log 2$ is only obtained if the following condition is met \cite{hastings}. Consider an annulus, such as that formed by the region $AB$ in Fig. \ref{fig1} (a). $\gamma_{LW} = \gamma_I = 2 \log 2$ if, by local operations on the spins within the annulus, a state can be prepared such that the parity of anyons on both the plaquettes and vertices enclosed by the the annulus (i.e. those with either full or partial support on spins within the region enclosed by the annulus)  is in a definite state. 

Let us now consider the following modified version of the toric code Hamiltonian,
\be \nonumber
H_1 = -J \sum_{\langle v,v' \rangle} A_v A_{v'} -J \sum_{\langle p,p' \rangle} B_p B_{p'}.
\ee
Here $\langle v,v' \rangle$ denotes a pair of vertices joined by an edge, and $\langle p,p' \rangle$ denotes a pair of plaquettes which share an edge. Hence, rather than single vertex and single plaquette terms of $H_0$, the Hamiltonian is composed of nearest neighbour two-vertex and two-plaquette interactions. Nevertheless, it is important to note that it is still both gapped and local, and has the same eigenstates as $H_0$. The ground states of $H_1$ correspond to the states for which the either no vertices hold an anyon, or all vertices hold an anyon. The same is true for the plaquettes. Let us consider the equally weighted mixture of all ground states, $\rho^{(1)}$.

In this case, the natural choice for the operator $O_i$ for the correlation function $G_{i,j}$ is $A_v$ (alternatively, $B_p$). It can then be easily seen that $G_{v,v'}(\rho^{(1)}) = 1$ for any pair of vertices $v$ and $v'$, without any decay due to the distance between them. The state therefore effectively has an infinite correlation length, despite being composed of ground states of a gapped and local Hamiltonian.

It is clear that this state is topologically ordered, with $\gamma_{LW} = \gamma_I = 2 \log 2$, since measurement of even a single vertex and plaquette on an annulus is sufficient to determine the occupancies of all vertices and plaquettes. The parity of both vertex and plaquette occupations enclosed by the annulus is then both definite and known.

To perform the calculation of the topological mutual information, note that state $\rho^{(1)}$ can be decomposed as an equally weighted mixture of four mixed states. Each of these corresponds to a different anyon configuration (vertices and plaquettes all empty, vertices filled and plaquettes empty, vertices empty and plaquettes filled, and vertices and plaquettes all filled), and hence they are orthogonal. Also, the entropies for each of these mixed states are equal, since they are equivalent by local unitaries. The reduced density matrices $\rho^{(1)}_R$ and $\rho^{(1)}_{R_c}$ are similarly composed of four orthogonal and locally unitarily equivalent mixed states, since the different anyon configurations can also be distinguished and manipulated within each region. The behaviour of the Von Neumann entropy when applied to block diagonal matrices \cite{iblisdir} then means that the mutual information can be decomposed into two contributions. The first comes from the correlations within each of the orthogonal and locally equivalent states. Since these states are all equivalent to $\rho^{(0)}$, this contribution is $I_{R}(\rho^{(0)})$. The second contribution comes from the fact that there are four states, distinguishable in both regions, that are equally mixed. This adds an amount $2 \log 2$ to the total mutual information. Hence,
\be \label{decomp}
I_{R}(\rho^{(1)}) = I_{R,R_c}(\rho^{(0)}) + 2 \log 2.
\ee
When these mutual informations are combined according to Eq. (\ref{KP}), the contribution of $I_{R}(\rho^{(0)})$ from each will lead to a total contribution of $2 \log 2$. This is due to the topological correlations within the state $\rho^{(0)}$. However, the total contribution of the $2 \log 2$ terms will be $-2 \log 2$. In total, therefore, the calculation for the topological mutual information yields $\gamma_I+\ldots=0$.

This result may lead anyone calculating this quantity to believe that $\rho^{(1)}$ does not possess topological correlations, and therefore that the nearest neighbour interactions of $H_1$ are not sufficient to support topologically ordered states. However, such a conclusion would be incorrect. The result of $\gamma+\ldots=0$ arises in this case only because non-topological correlations are not suppressed, no matter how large the regions are made, and are therefore able to cause ambiguity.

In general, any state on a 2D surface for which the spins (or other physical medium) around one point have correlations with other those around points spread throughout the surface, and the correlations do not decay with distance, will cause ambiguities in the calculation of the topological mutual information. If pure states occur with the same properties, similar ambiguities will arise for the topological entanglement entropy that it is based upon \cite{kitpres}. This is because such correlations cause the mutual information across any bipartition to have contributions that do not depend on the length of the boundary, and hence cannot be distinguished from genuine topological contributions.

For the second example we consider another modified toric code Hamiltonian,
\be \nonumber
H_2 = -J \sum_{( v,v' )} A_v A_{v'} -J \sum_{( p,p' )} B_p B_{p'}.
\ee
Here $(v,v')$ denotes a pair of vertices joined by a vertical edge, and $(p,p')$ denotes a pair of plaquettes which share a horizontal edge. As such, $H_2$ is like $H_1$ in that it consists of nearest neighbour 2-vertex and 2-plaquette interactions. But, unlike $H_1$, these interactions are confined to a specific direction. The ground state of $H_2$ consist of any state for which each vertical line of vertices or plaquettes is either empty or full of anyons. We consider the equally weighted state $\rho^{(2)}$ of all ground states.

As before, using $O_i = A_v$ yields $G_{v,v'}(\rho^{(2)}) = 1$ for any pair of vertices $v$ and $v'$ on the same line. There is again no decay due to distance, and hence an infinite correlation length despite being composed of ground states of the a gapped and local Hamiltonian. 

This state is again topologically ordered with $\gamma_{LW} = \gamma_I = 2 \log 2$. This is because measurement of a vertex (plaquette) on each line of vertices (plaquettes) that intersect the region enclosed by the annulus is sufficient to determine the parity of the vertex (plaquette) occupations within this region. 

Like $\rho^{(1)}$, the state $\rho^{(2)}$ is a sum of orthogonal and locally equivalent states of different anyon configurations, with each state locally equivalent to $\rho^{(0)}$. The same is true for the reduced states for any region. The entropy of a region $R$ is therefore composed of an $S_R(\rho^{(0)})$ contribution due to the correlations within each state and a contribution of $\log 2$ for each vertical line of plaquettes or vertices that intersects the region (i.e. has at least one vertex or plaquette with full support on the region). If we use $n_R$ to denote the number of such lines, the mutual information between two regions is
\be \label{decomp}
I_{R,R'}(\rho^{(2)}) = I_{R,R'}(\rho^{(0)}) + (n_R + n_{R'} - n_{RR'}) \log 2.
\ee
When these are combined according to Eq.(\ref{LW}), the result will be $\gamma_{LW}+\ldots= 2 \log 2 + n \log 2$. Here $n$ denotes the number of vertical lines that intersect both $A$ and $B_2$ but do not intersect $B_1$. The value of the topological entropy is therefore overestimated by an amount that is not suppressed as the size of the regions are increased, but in fact increases. This is due to the non-topological correlations along lines of anyons. The same is true for the reformulation of the Levin and Wen method in \cite{claudios1}.

In general, any states with infinite range non-topological correlations along lines are able to cause ambiguities to the topological entropy of Levin and Wen. This is because such correlations are able to directly correlate the regions $A$ and $B_2$, and the calculation is built on the assumption that such correlations do not occur.

With these examples, we see that both of these topological order parameters are susceptible to long range non-topological correlations, since they can cause corrections that cannot be suppressed. This may then lead to non-topological phases being identified as topological by mistake, or vice versa, especially as more exotic systems are probed \cite{zang,zheng,sen,son}. Though both examples considered only fool one of the parameters, with $\rho^{(1)}$ and $\rho^{(2)}$ both giving the expected values of $2 \ln 2$ when using $\gamma_{LW}$ and $\gamma_I$ respectively, clearly states will exist which fool them both. As such, a way to rid the entropy of the non-topological corrections must be determined.

\section{Modifying the topological entropy}

To solve the problem, let us take a closer look at how the topological entropy is defined. In Eq. (\ref{LW}) it was shown that $\gamma_{LW}$ is the information that a section of a closed string ($A$) shares with the rest of the string ($B$), but not with the neighbouring sections ($B_1$). Since it can usually be expected that any contributions to $I_{A,B}$ due to local correlations act over the boundary between $A$ and its neighbour $B_1$, this definition serves to remove all local boundary correlations and thus, in most cases, measures only the topological correlations around the string. The problem, as mentioned above, is that sufficiently long range local correlations contribute in a way that does not depend on the boundary, and can correlate the non-neighbouring regions $A$ and $B_2$ directly. Solving the problem therefore requires modifying the calculation of the topological entropy such that local correlations do not cause unsuppressable corrections, whatever their range. To do this let us define the topological entropy $\gamma$ as the information that the region $A$ shares with $B$, but not $B_1$ or $B_2$ individually. 

This definition isolates the correlations that can be seen only when all the three regions of the annulus are considered. Only correlations that contribute to $I_{A,B}$, but not to $I_{A,B_1}$, $I_{A,B_2}$ or $I_{B_1,B_2}$ will contribute to $\gamma$. Since non-topological correlations characterized by two-point correlation function will indeed contribute to the latter three mutual informations, they cannot affect the value of $\gamma$, and so cannot cause ambiguity whatever their range. Only truly non-local correlations, such as those existing along strings in topological ordered states, will therefore remain. Note that this definition only requires that the three regions together form an annulus, and the composite of any two regions does not. As such partitioning the closed loop according to Fig. \ref{fig1}(b) is therefore no longer a requirement (though it is still valid, and we will continue to use in this study), and other tripartions of an annulus may be used.

The only caveat is as follows. Suppose that infinite range non-topological correlations are present in a state such that the operators $O_i$ are not defined on a single spin $i$, but of spins within its neighbourhood also. Consider then a point $j$ that lies on the boundary between the regions $B_1$ and $B_2$, such that the support of $O_j$ is split between these two regions, and consider also a point $i$ that lies fully within the region $A$. Any two point correlations between $i$ and $j$ can therefore only be revealed when all three regions are considered. Such correlations would therefore appear to contribute to $I_{A,B}$ without contributing to $I_{A,B_1}$, $I_{A,B_2}$ or $I_{B_1,B_2}$, and so cause ambiguities with the value of $\gamma$. However, assuming the Hamiltonian causing the correlations is local, the points $i$ and $j$ will not be correlated in isolation. There will be other points $k$ with which $i$ shares exactly the same information it shares with $j$. In the large annulus limit, at least some of these will exist such that that $O_k$ has full support within $B_1$ or $B_2$. The contribution to $I_{A,B}$ will also be present in $I_{A,B_1}$ or $I_{A,B_2}$, respectively, and so be removed from the calculation along with all other non-topological contributions. 

Calculation of the modified topological entropy is not trivial, even when only classically correlated states are considered. Any calculation must distinguish the information that $A$ shares with both $B_1$ and $B_2$, and that which it shares only with the composite region $B$. In order to make the distinction, and calculate the modified topological entropy, we can use the state of the system, $\rho$ to define an alternative state $\rho'$. This is constructed such that $\rho_{AB_1} = \rho'_{AB_1}$ and $\rho_{AB_2} = \rho'_{AB_2}$. The correlations between the regions $A$ and $B_1$ are therefore exactly the same for $\rho'$ as they are for $\rho$, as are the correlations between the regions $A$ and $B_2$. We then maximize the entropy $S(\rho'_{AB})$ over all states for which these conditions hold. Any additional correlations that $A$ has only with the composite region $B$ will then not occur in $\rho'$. The topological entropy is then,
\be
 \gamma = I_{A,B}(\rho)-I_{A,B}(\rho').
\ee
This calculation involves both reconstruction of a joint state from reduced density operators and a maximization. As such, computation of this quantity may be difficult in general. A simple method of calculating $\gamma$ in this way does exist if the states are classical, i.e. $\rho$ takes the form $\rho = \sum_{a,b_1,b_2} p(a,b_1,b_2) \ket{a,b_1,b_2}\bra{a,b_1,b_2}$, where $\{\ket{a}\}$ is a basis for the states of $A$, etc. In this case the state $\rho'$ can then be defined using the probabilities $p'(a,b_1,b_2) = p(a)p(b_1|a) p(b_2|a)$, the product of the marginal distributions for $B_1$ and $B_2$ conditioned on $A$. This maintains the correlations between $A$ and $B_1$, and between $A$ and $B_2$, but breaks those between $A$ and $B$ alone.

Even without an efficient means to calculate $ \gamma$ in general, both upper and lower bounds may easily be determined, applicable in both the quantum and classical cases. To calculate these bounds, we consider the quantity $I_{A,B}- \gamma$, the information $A$ shares with $B_1$ and $B_2$ individually but not with $B$ alone. This satisfies the bounds,
\bq\nonumber
I_{A,B}-  \gamma &\geq& \max(I_{A,B_1},I_{A,B_2}), \\ \nonumber
I_{A,B}-  \gamma &\leq& I_{A,B_1}+I_{A,B_2}.
\eq
The first of these come from the fact that $I_{A,B}-  \gamma$ is at least the information $A$ shares with $B_1$, or that which it shares with $B_2$. Taking the maximum of the two hence forms a lower bound. The second comes from the fact that adding the information $A$ shares with $B_1$ to that which it shares with $B_2$ will count all of the information in $I_{A,B}-  \gamma$, and may even double count some, and so forms an upper bound. Using these, and noting that $\gamma$ cannot be less than zero, it follows that,
\bq \label{new} \nonumber
  \gamma &\geq& \min(I_{A,B} -I_{A,B_1}-I_{A,B_2},0), \\
  \gamma &\leq& I_{A,B} - \max(I_{A,B_1},I_{A,B_2}).
\eq
This gives bounds for $\gamma$ in terms of the mutual informations for bipartitions. Note that these are exactly the kinds of quantities used to evaluate Eq. (\ref{KP}) and Eq. (\ref{LW}) above. As such, computation of these bounds is no harder than computation of existing quantities.

To demonstrate that the modified topological entropy correctly identifies topologically ordered states, the bounds may be calculated for the string-net models \cite{stringnet}, using the relations from \cite{levwen}. For these, the entropy of a region $R$ is given by,
\be \label{topo}
S_R = - j_R \log D - n_{R} \sum_a \frac{d_a^2}{D} \log \left( \frac{d_a}{D} \right),
\ee
where $j_R$ is the number of disconnected boundary curves in $R$ and $n_{R}$ is the number of spins along the boundary (see \cite{levwen} for more details on the specific models considered). The two regions $A$ and $B_2$ are not neighbouring, and so $j_{AB_2'}=j_A+j_{B_2}$ and $n_{AB_2}=n_A+n_{B_2}$. As such, we find that $I_{A,B_2}=0$, and so the regions are not correlated. The upper and lower bounds of Eq. (\ref{new}) then coincide both with each other, and with the definition of $\gamma$ in Eq. (\ref{LW}). This gives the topological entropy a unique value for these states, equal to the expected result of $  \gamma=2\log D$. 

For the state $\rho^{(1)}$, $I_{R,R'}(\rho^{(1)}) = I_{R,R'}(\rho^{(0)}) + 2 \log 2$. As such, since we know that the topological correlations result in $I_{A,B}(\rho^{(0)}) - I_{A,B_1}(\rho^{(0)}) = 2 \log 2$, we see that the lower bound of Eq.(\ref{new}) vanishes and the upper bound gives $2 \log 2$. For the state $\rho^{(2)}$, the lower bound vanishes and the upper bound diverges with increasing region size.

The bounds for $\rho^{(1)}$ and $\rho^{(2)}$ therefore do not assign a unique value to $\gamma$ for in either case, nor can they tell us whether $\gamma=0$ or $\gamma>0$. Nevertheless, the fact that they are not equal is a clear signature of long-range non-topological correlations. Once their presence has been detected, efforts to remove the ambiguity they cause can be made. This is in contrast to previous methods, which assign a single value and give no clue as to whether this stems from topological correlations, or long range non-topological correlations.

\section{Conclusions}

Here we have shown that existing definitions of the topological entropy and topological mutual information can give misleading results, since they cannot completely distinguish between topological and non-topological correlations in all cases. As such, a modified definition of the topological entropy, $ \gamma$, has been proposed. This is equivalent to the topological entropy of Levin and Wen in most cases. However, unlike existing definitions, $ \gamma$ is robust against all non-topological correlations, even if they effectively have an infinite correlation length. Upper and lower bounds on this modified entropy have been determined, which can be computed with the same complexity required for previous methods. These bounds will coincide in most cases and so provide a unique value for $ \gamma$. In the the presence of long range non-topological correlations, the bounds diverge. This serves as a witness to the presence of such correlations, as well as limiting the ambiguity they can cause.

It should be noted that problems arising from infinite correlation lengths do not occur only for mixed states. Pure eigenstates of gapless Hamiltonians will similarly cause ambiguities in the calculation of the topological entropy. Though a full treatment of gapless states and the ambiguities they cause to the topological entropy is beyond the scope of this work, it seems likely that the modified topological entropy defined here will be useful also in the gapless case.

\section{Acknowledgements}

The author would like to thank Alioscia Hamma for critical reading of the manuscript and useful discussions, and especially for pointing out that non-topological correlations have a cancellation effect on $\gamma_I$. Thanks also to Jiannis K. Pachos and Abbas Al-Shimary for critical reading of the manuscript and useful discussions.


\begin{thebibliography}{99}
\bibitem{note1} Equivalent definitions of the regions could be used for which this is not true for the state considered. However, there is always a way to define the lines to give an equivalent result for any choice of regions.
\end{thebibliography}
\end{document}